%% file: proceedings.tex
\documentclass{PoS}

\title{Determination of the $\Delta$ resonance width from lattice QCD}

\ShortTitle{$\Delta$ resonance from lattice QCD}

\author{Constantia Alexandrou\\
    Department of Physics, University of Cyprus, P.O. Box 20537, 1678 Nicosia, Cyprus, and\\  
 Computation-based Science and Technology Research  
    Center, Cyprus Institute, 20 Kavafi Str., Nicosia 2121, Cyprus \\
        E-mail: \email{c.alexandrou@cyi.ac.cy}}

\author{John W. Negele\\
Center for Theoretical Physics, Laboratory for Nuclear Science and 
        Department of Physics, Massachusetts Institute of Technology, Cambridge, Massachusetts 02139, U.S.A.\\
        E-mail: \email{negele@MIT.EDU}}

\author{\speaker{Marcus Petschlies}\\
 Computation-based Science and Technology Research  
    Center, Cyprus Institute, 20 Kavafi Str., Nicosia 2121, Cyprus \\
        E-mail: \email{m.petschlies@cyi.ac.cy}}

\abstract{
  A method suitable for extracting resonance  parameters of unstable
  baryons in lattice QCD is examined.  The method is applied 
  to the strong decay of the $\Delta$ 
  to a pion-nucleon state, extracting the $\pi N\Delta$ coupling constant and $\Delta$ decay width.
}

\FullConference{31st International Symposium on Lattice Field Theory - LATTICE 2013\\
		July 29 - August 3, 2013\\
		Mainz, Germany}

\input{adjustment.tex}

\begin{document}

\section{Introduction}

The study of hadron resonances is 
of fundamental importance for nuclear and particle physics and one of the long-standing
goals of lattice QCD. While methods for handling stable low-lying particles on the lattice
are well developed, the study of resonances and decays is intrinsically more difficult within the Euclidean formulation of  lattice QCD.
The problem lies essentially in the fact, that scattering states cannot be realized for a theory formulated in
Euclidean time.

Several ways to resolve the issue of relating data extracted from finite volume states in lattice QCD
and properties of scattering states in infinite volume have been proposed.
The method introduced by L{\"u}scher relates the shifts in the energy spectrum of multi-hadron states
due to their interaction in finite volume to the scattering phase shifts in infinite volume~\cite{Luscher:1985dn,Luscher:1986pf}.
In its full-fledged version this requires lattice simulations with a series of spatial lattice volumes with precise
determinations of the spectrum of multi-hadron states. This method has been applied mainly to meson resonances
\cite{Torok:2009dg,Feng:2010es,Li:2007ey} and also to the negative parity nucleon channel~\cite{Lang:2012db}.

An alternative approach to study strong decays on the lattice
has been proposed in Ref.~\cite{McNeile:2002az}. While a true resonant behavior cannot appear
for finite-volume states on a Euclidean lattice, initial and final states , $\left|\,i\,\right.\rangle$,
$\left|\,f\,\right.\rangle$, with matching quantum numbers
can mix, meaning that states realized on the lattice at asymptotically large times are a linear  combination of
 states of the same quantum numbers. This mixing will be 
 significant for states with equal energy.
The so-called transfer matrix method provide a method to  extract the overlap  amplitude  of such states 
and relate it to the leading-order continuum matrix element 
$\matelem\sim \brackets{f\,\left|\,H\,\right|i}$.
The matrix element is then related to the resonance parameters for the transition, namely the coupling and the width.

In this study we apply the transfer matrix method to baryonic resonances and specifically to the transition $\Delta \to \pi N$, which has been  well-known studied experimentally. Thus, this work aims at benchmarking the method for an experimentally known
width before applying it to other resonances where the width is either poorly determined or not yet measured. The methodology and some of the results were reported in
Ref.~\cite{Alexandrou:2013ata}.

\section{Transfer matrix method}

\begin{figure}[htpb]
  \begin{center}
    \includegraphics[width=0.8\textwidth]{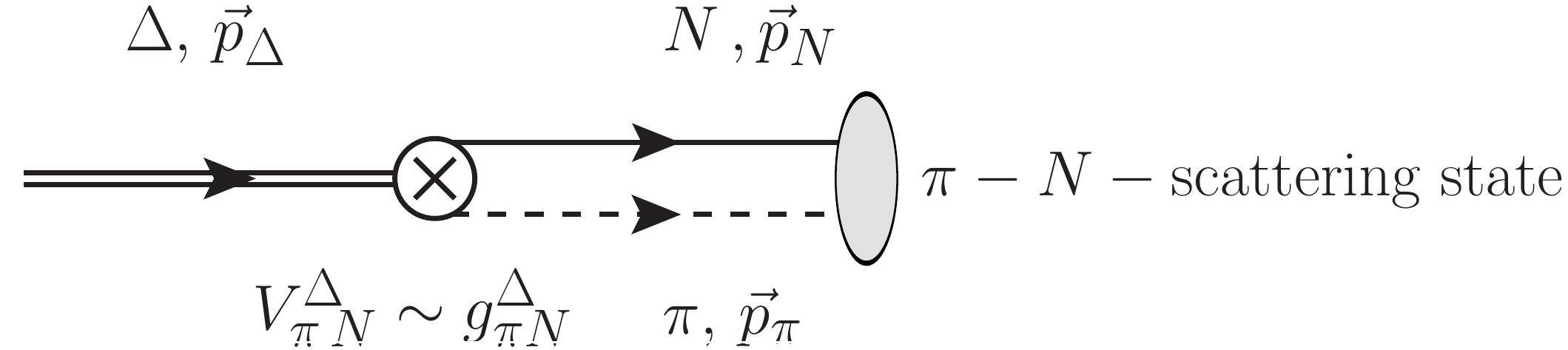}
  \end{center}
  \caption{Schematic presentation of the $\Delta$ resonance and its decay channel,
    the pion-nucleon  with coupling $g_{\pi N}^\Delta$.
  }
  \label{fig:1}
\end{figure}

We consider the $\Delta\rightarrow \pi N$ transition depicted in Fig.~\ref{fig:1}, as described in an effective field theory approach, where  an initial $\Delta$ state  couples resonantly to the $\pi N$ state.
  We associate this interaction with a vertex
described by an effective coupling $g^\Delta_{\pi N}$. The transfer matrix
in the subspace  $\left\{|\Delta\rangle,\, |\pi\,N\rangle\right\}$ spanned by the pion-nucleon and
the $\Delta$ states, which without an interaction are orthogonal, can be written as 
\begin{eqnarray}
  \mathrm{T} &=&  \epow{-a\bar{E}}\,
    \begin{pmatrix}
      \epow{-a\delta/2} & ax                & \cdots \\ 
      ax                & \epow{+a\delta/2} &        \\
      \vdots            &                   & \ddots \\
    \end{pmatrix}
    \label{eq:1}
\end{eqnarray}
where we denote by $\bar{E} = \rbrackets{ E_\Delta + E_{\pi N}}/2$ the average energy of the $\Delta$ and  $\pi N$ system and by $\delta = E_{\pi N} - E_\Delta$  the energy difference between the pion-nucleon and the 
$\Delta$ state. The transition amplitude
is denotes by $x = \brackets{\Delta\,\left| \,H\,\right|\pi N}$.

The coupling of the $\Delta$ to the $\pi N$ state  causes an energy  shift 
of the energy eigenstates of the  non-interacting theory. The energies in the
 the subspace $\left\{|\Delta\rangle,\, |\pi\,N\rangle\right\}$ are modify to 
\begin{eqnarray} 
  E_\pm &=& \bar{E} \pm \sqrt{\delta^2/4 + x^2}\,.
  \label{eq:2}
\end{eqnarray}

On the lattice we prepare the pion-nucleon state at an initial time $t_i$ and the $\Delta$ state at a final time $t_f$ and sum 
over all possibilities for one transition (leading-order) at each intermediate time. The resulting transition overlap amplitude is given by 
\begin{eqnarray}
  \brackets{\Delta,\,t_f\,|\,\pi N,\,t_i} &=& 
    ax\,\frac{\sinh(\delta\,t / 2)}{\sinh(a\delta/2)}\,\epow{-\bar{E} t}
    \xrightarrow[]{\delta t \ll 1}  \sbrackets{ax\,t}\,\epow{-\bar{E} t} + \ldots\,,
  \label{eq:3}
\end{eqnarray} 
where $t \equiv t_f-f_i= a n_{fi}$ and $n_{if}$ is the number of transition steps.
Higher order contributions, contributions from excited states
 and mixing with other states are denoted by the terms omitted in left-hand-side of Eq.~(\ref{eq:3}).

For the method to be applicable  the following  conditions must hold:
\begin{enumerate}
  \item the energy levels of the initial and final states must be sufficiently close 
  \item the transition amplitude needs to be sufficiently small so that
only one transition occurs  (leading-order
    contribution)
\end{enumerate}
 Since we need to have final states with non-zero spatial momentum on the lattice, with
    a sufficiently fine resolution in momentum space, a large enough spatial volume is preferable.

\begin{figure}[htpb]
  \begin{center}
    \includegraphics[width=0.75\textwidth]{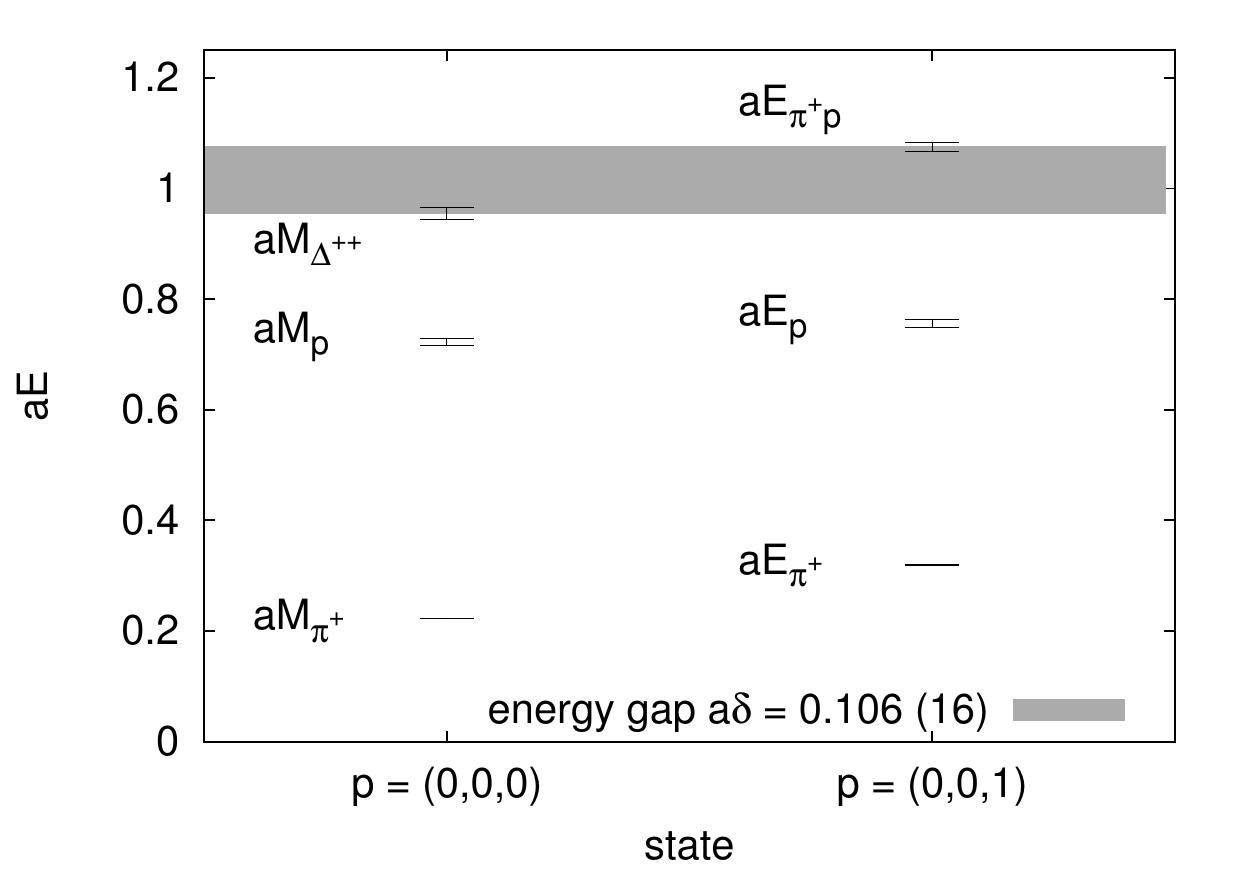}
  \end{center}
  \caption{The  relevant energy levels for the ensemble of  domain wall valence
    quarks on a staggered sea for $\mpi = 360\mev$. We show the single
    particle masses $M_{\pi^+}$, $M_p$ and $M_{\Delta^{++}}$ and the
    sum of the pion and proton energy $E_{\pi^+ p} = E_{\pi^+} + E_p$. 
  }
  \label{fig:2}
\end{figure}

In our numerical study, we consider the $I_z=3/2$ channel with  an initial $\Delta^{++}$ and a final $\pi^+$-proton state. We illustrate how well the first condition  on the energy levels 
is satisfied for our current study  in Fig.~\ref{fig:2}, where we show the computed energy of the $\Delta^{++}$ and the sum of the energies of the $\pi^+$ and 
the proton $E_{\pi^+ p}=E_{\pi^+}+E_p$.

\section{Lattice calculation}
In the lattice calculation the unknown overlaps and exponential time dependence cancel by taking a suitable ratio of 3-point and 2-point functions 
\begin{eqnarray}
  R_\mu( t,\Qvec,\qvec) &=& \frac{ C^{\Delta \to \pi N}_\mu( t,\,\Qvec,\qvec) }{
    \sqrt{ C^{\Delta}_\mu( t,\,\Qvec) \,C^{\pi N}( t,\,\Qvec,\,\qvec)}}\,.
  \label{eq:6}
\end{eqnarray}
As already pointed out, we consider the isospin $I_z = +3/2$ channel and use the standard interpolating fields for the $\Delta^{++}$, the $\pi^+$ and the proton.
 The $\pi^+ p$ state is represented by the interpolating operator
\begin{eqnarray}
  J_{\pi^+ p}^\alpha(t,\qvec,\xvec) &=& \sum\limits_{\yvec}\,J_{\pi^+}(t,\yvec+\xvec)\,J_p^\alpha(t,\xvec)\,\epow{-i\qvec \cdot \yvec}
  \label{eq:7}\,,
\end{eqnarray}
defined to have a relative momentum $\vec{q}\ne \vec{0}$ to generate overlap with the $\pi^+ p$ state with angular momentum
$l=1$. The dominant asymptotic contribution to the correlator will then come from the coupling
$s_p \oplus l \rightarrow J_\Delta = 3/2$.
Moreover, in this study we neglect the interaction between the pion and the
proton in a finite box and  calculate the pion-proton  2-point function,
as a product of the pion and proton correlator, i.e.
$C^{\pi^+ p}\approx C^{\pi^+}\times C^p$.

We use a  hybrid action of staggered sea quarks 
and domain wall (DW) valence quarks. The gauge field configurations are taken from the MILC ensemble \textit{2864f21b676m010m050}
\cite{Bernard:2001av,WalkerLoud:2008bp} with pion mass $\mpi \approx 360\mev$, lattice spacing $a \approx 0.124\fermi$ and  spatial volume of
$\rbrackets{3.4\fermi}^3$. We use 210 configurations with 4 measurements per configuration.
The relative momentum in the pion-proton system is set to $\qvec = (2\pi/L)\,\vec{e}_k$,
$k=\pm 1,\,\pm 2,\,\pm 3$ and we combine data from both all six momentum directions as well as forward and backward
propagation. 
\begin{figure}[htpb]
  \begin{center}
    \includegraphics[width=0.48\textwidth]{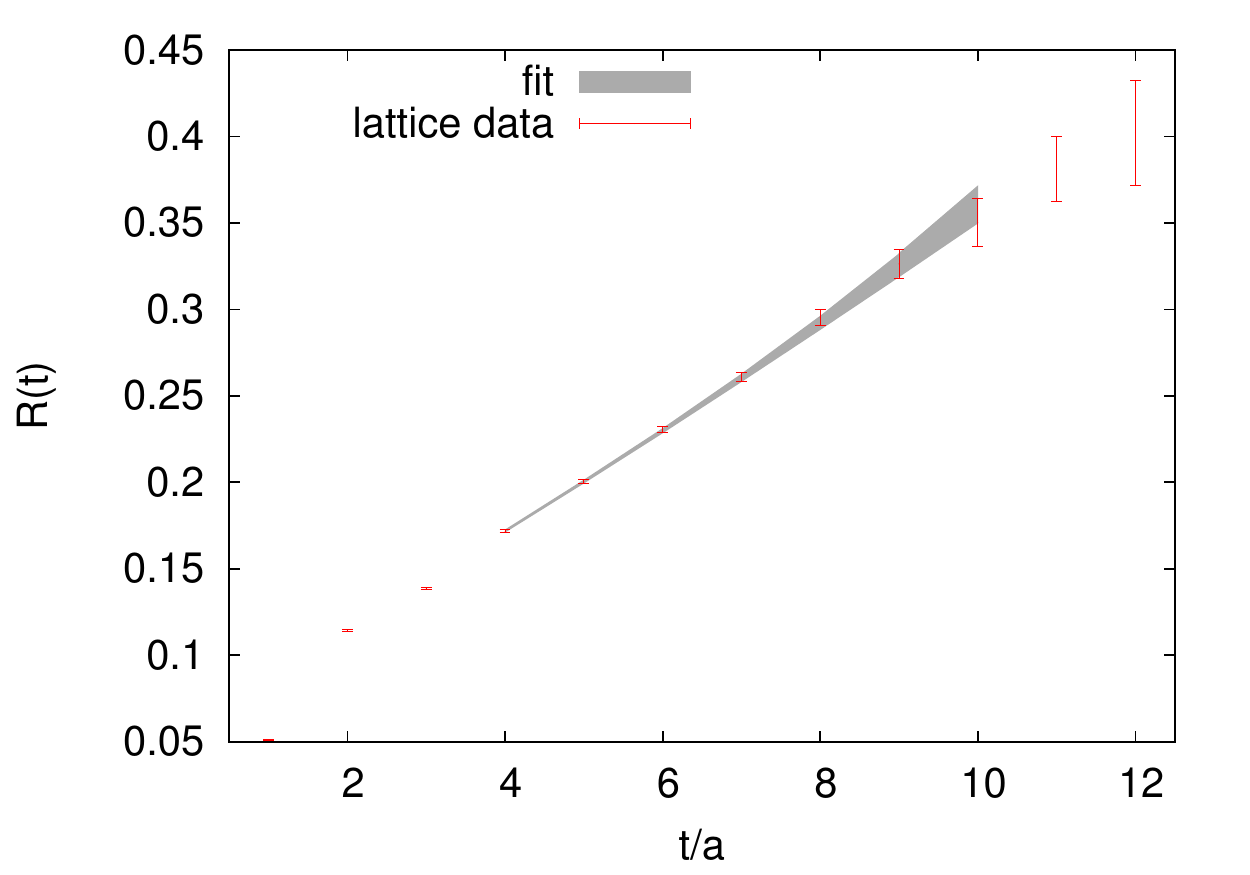}
    \includegraphics[width=0.48\textwidth]{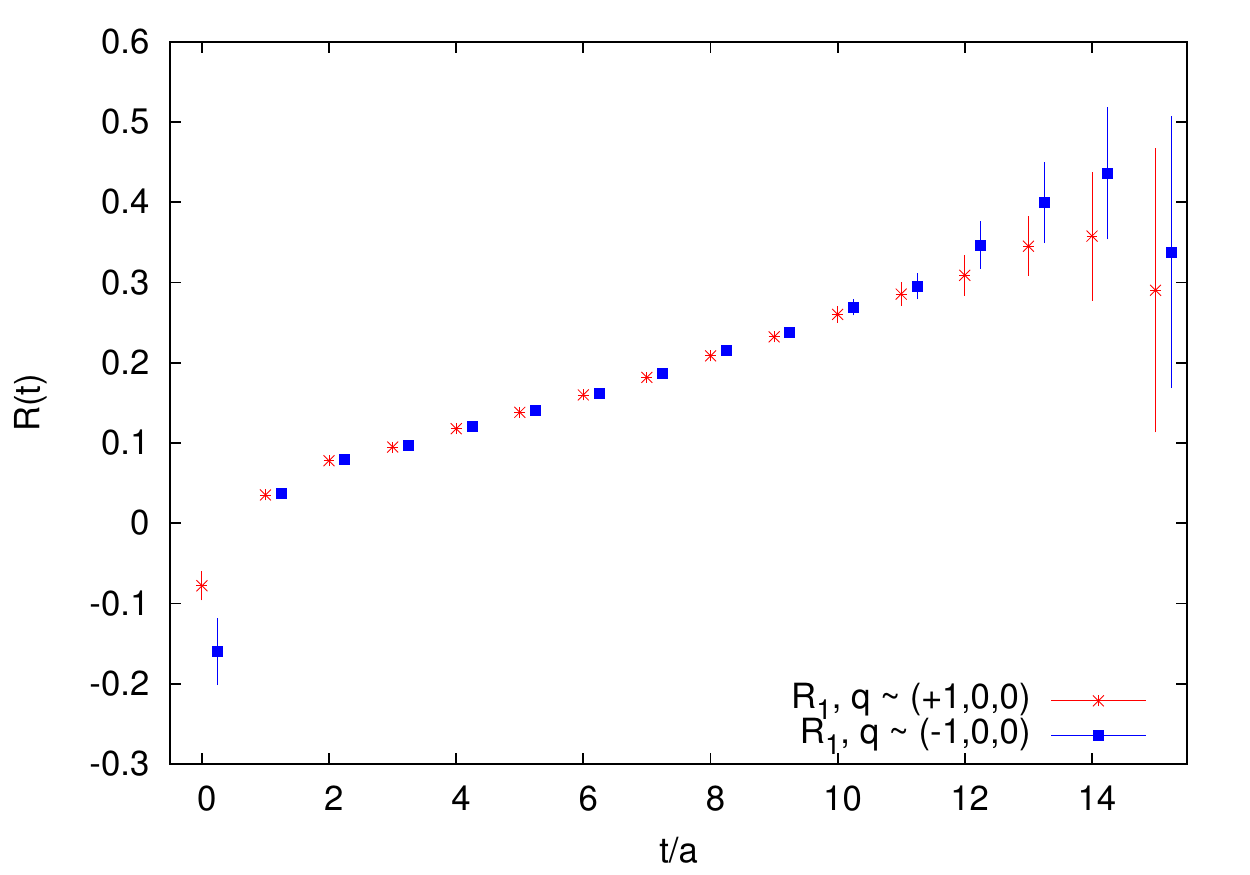}
  \end{center}
  \caption{Ratio $R$ averaged over over momentum directions and forward / backward propagation as measured in our
    numerical study.}
  \label{fig:4}
\end{figure} 
In the left panel of Fig.\ref{fig:4}, we show the ratio $R$ of Eq.~(\ref{eq:6}), which results
from averaging the six contributions $R_j$, $j=1,2,3$ with
$\qvec\propto \pm \vec{e}_j$.
If $ t x \ll 1$ then $R \propto t$.
As can be seen, 
the data
suggest the existence of  a time  interval  $4 \lesssim t/a \lesssim 10$ where R has a linear dependence on the sink-source separation $t$. We remark, that the signal on the
smallest timeslices is affected by excited states contamination, most notably by contributions from oscillatory
contributions known to exist for domain wall fermions~\cite{Syritsyn:2007mp}. At large time separations, the evaluation of the
ratio is essentially limited by the two-point function of the $\Delta$, which enters to the power $-1/2$ and becomes statistically compatible
with zero for $t \gtrsim T/4$.

To extract the overlap $x$ we fit the time-dependence of $R$ using two Anse\"aze, namely
\begin{equation}
  f_1(t) = A + B\,a\,\frac{\sinh(\delta\,t/2)}{\sin(a\delta/2)}\,,\hspace*{1cm}
  f_2(t) = A + B\,t \: ( + C\,t^3 )\,.
  \label{eq:8}
\end{equation}
$f_1$ is the form we expect from the transfer matrix analysis as given in  Eq.~(\ref{eq:3}), amended by an additive constant shift that
incorporates excited state contributions.
We expect the second Ansatz $f_2$ to be valid in the time window where $R$
is approximated by a  linear-dependence on time.
 The stability of the extracted slope $B$ is checked by including a cubic term  in $t$. The fit results
for a number of choices of fit windows $[t_\mathrm{min},\,t_\mathrm{max}]$ are compiled in Tab.~\ref{tab:1}.
\begin{table}
  \centering
    \begin{tabular}{ccccccc}
      & $t_\mathrm{min}/a$ & $t_\mathrm{max}/a$ & $A\cdot 10^{2}$ & $aB\cdot 10^2$ & $a^3C\cdot 10^5$ & $\chi^2/\mathrm{dof}$ \\
      \hline
      $f_1$ &  $4$ & $ 9$ & $6.47 \,( 49 )$ & $ 2.62 \,(15 ) $ & $ 0.188 \,( 68 )$ & $ 2.4 / 3$\\
      $f_1$ &  $4$ & $10$ & $6.24 \,( 47 )$ & $ 2.69 \,(14 ) $ & $ 0.156 \,( 79 )$ & $ 4.3 / 4$\\
      $f_1$ &  $5$ & $ 9$ & $5.62 \,(103 )$ & $ 2.82 \,(26 ) $ & $ 0.140 \,(104 )$ & $ 1.8 / 2$\\
      $f_1$ &  $5$ & $10$ & $5.05 \,( 84 )$ & $ 2.98 \,(21 ) $ & $ 0.074 \,(122 )$ & $ 2.9 / 3$\\
      \hline
      $f_2$ &  $4$ & $ 9$ & $5.62 \,(25 )$ & $ 2.89 \,(06 ) $ & & $ 6.0 / 4$\\
      $f_2$ &  $4$ & $10$ & $5.63 \,(25 )$ & $ 2.89 \,(06 ) $ & & $ 6.5 / 5$\\
      $f_2$ &  $5$ & $ 9$ & $4.75 \,(51 )$ & $ 3.05 \,(10 ) $ & & $ 2.4 / 3$\\
      $f_2$ &  $5$ & $10$ & $4.78 \,(52 )$ & $ 3.05 \,(11 ) $ & & $ 3.0 / 4$\\ 
      \hline
      $f_2$ &  $4$ & $ 9$ & $6.51 \,( 53 )$ & $ 2.60 \,(16 )$ & $ 4.1 \,(22 )$ & $ 2.4 / 3$\\
      $f_2$ &  $4$ & $10$ & $6.27 \,( 52 )$ & $ 2.68 \,(16 )$ & $ 2.9 \,(21 )$ & $ 4.3 / 4$\\
      $f_2$ &  $5$ & $ 9$ & $5.64 \,(128 )$ & $ 2.82 \,(33 )$ & $ 2.4 \,(32 )$ & $ 1.8 / 2$\\
      $f_2$ &  $5$ & $10$ & $5.05 \,(117 )$ & $ 2.98 \,(30 )$ & $ 0.7 \,(28 )$ & $ 2.9 / 3$\\
      \hline\hline
    \end{tabular}
  \caption{Results for the fit parameters $A$, $B$ and $C$  for the fit functions $f_1$ and $f_2$  and several choices of the fit interval.}
  \label{tab:1}
\end{table}

\section{Extraction of the coupling}
To extract the effective coupling from the parameter $B$ we connect to leading-order (LO) effective field theory (EFT) and find
\begin{eqnarray}
  B= \sum\limits_{\sigma_3,\,\tau_3}\,\frac{\matelem(\Qvec,\,\qvec,\,\sigma_3,\,\tau_3)}{\sqrt{N_\Delta\,N_{\pi N}}}\,V\,\delta_{\Qvec \Qvec} \times \mathrm{spin~sum~factor}\>.
  \label{eq:9}
\end{eqnarray}
The finite volume normalization of the states reads
\begin{equation}
  N_\Delta = V\,\frac{E_\Delta}{m_\Delta}\,,\hspace*{0.5cm}
  N_{\pi N} = N_\pi \times N_N = 2V\,E_\pi \times V\,\frac{E_N}{m_N}\,.
  \label{eq:10}
\end{equation}
The matrix element $\matelem$ is decomposed according to LO EFT as
\begin{eqnarray}
  \matelem(\Qvec,\,\qvec,\,\sigma_3,\,\tau_3) &=& \frac{g^{\Delta}_{\pi N}}{2m_N}\,\ubar_\Delta^{\mu\,\alpha}(\Qvec,\sigma_3)\,q_\mu\,u_N^{\alpha}(\Qvec+\qvec,\tau_3)\,.
  \label{eq:11}
\end{eqnarray}
As can be seen from  Eq.~(\ref{eq:11}) for $\qvec \propto \vec{e}_j$,
 the imaginary part of  $R_j$ is the only component that should have a non-zero signal. This serves as a consistency check of our results, which satisfy this expectation.
Using the values of $B$ given in Tab.~\ref{tab:1} we extract a the coupling
\begin{eqnarray}
  g_{\pi N}^\Delta\rbrackets{\mathrm{LAT}} = 27.0 \,\pm 0.6\,\pm 1.5\,.
  \label{eq:12}
\end{eqnarray}
The second error is a systematic error estimated from the spread of the results due to the  different fit intervals and fit functions.
To compare this result on the coupling constant with the experimental value we relate the
coupling to the width,
\begin{eqnarray}
  \Gamma &=& \frac{ {g_{\pi N}^{\Delta}}^2}{48\pi}\,\frac{1}{m_N^2}\,\frac{E_N + m_N}{E_N + E_{\pi}}\,q^3\,.
  \label{eq:14}
\end{eqnarray}
Using the PDG value for the $\Delta$ width~\cite{Beringer:1900zz} we find
$g_{\pi N}^\Delta\rbrackets{\mathrm{LO~EFT}} = 29.4\,(4)$.
In Ref.~\cite{Hemmert:1994ky} a model-independent K-matrix analysis yielded the value
$g_{\pi N}^\Delta\rbrackets{\mathrm{EXP}} = 28.6\,(3)$.
We find reasonable agreement with both results. This, in turn, means that the 
width 
\begin{eqnarray}
  \Gamma_\Delta\rbrackets{\mathrm{LAT}} &=& 99\,(12)\mev\,,
  \label{eq:15}
\end{eqnarray}
obtained using our lattice result on the coupling constant is consistent with the experimental value
$\Gamma_\Delta\rbrackets{\mathrm{EXP}} = 117\,(3)\mev$.

\section{Discussion and Outlook}
We presented results on the $\Delta$ resonance 
for  one ensemble of staggered sea quarks and domain wall valence quarks.
Although we find good agreement with the experimental value one needs to investigate lattice artifacts as well as perform the computation with smaller pion mass before a final result can be given. However, this study has shown   that the method
yields robust results and therefore, one can apply it  
to study the decay of other baryons. As an outlook to future work we show
preliminary results for the analysis of the decay $\Sigma^{*+} \to \pi^+ \Lambda^0$ in the right panel of Fig.~\ref{fig:4}.
Using only momentum directions $\qvec \propto \pm \vec{e}_1$ we obtain the preliminary results
\begin{eqnarray}
  4 \le t/a \le 12 \quad aB &= 0.0208\,(06) \mathrm{~with~} \chi^2/\mathrm{dof} = 1.1 \Rightarrow g = 21.5 \pm 0.7\nonumber\\
  6 \le t/a \le 12 \quad aB &= 0.0228\,(16) \mathrm{~with~} \chi^2/\mathrm{dof} = 1.1 \Rightarrow g = 23.6 \pm 1.6\,,
  \label{eq:15}
\end{eqnarray}
which can be compared to the LO EFT result extracted from  the width \cite{Beringer:1900zz},
\begin{eqnarray}
  g^{\Sigma^{*+}}_{\pi \Lambda}\rbrackets{\mathrm{LO~EFT}} \approx 20.0
  \label{eq:16}
\end{eqnarray}

\noindent
{\bf Acknowledgments:} We thank C. Michael for valuable discussions.
This research was in part supported by the Research Executive Agency of the 
European Union under Grant Agreement number PITN-GA-2009-238353 (ITN STRONGnet)
and in part by the DOE Office of Nuclear Physics under grant \#DE-FG02-94ER40818.
Computing resources were provided by the Cyprus Institute supported in
part by the Cyprus Research Promotion Foundation under contract
NEA Y$\Pi$O$\Delta$OMH/$\Sigma$TPATH/0308/31, the National Energy Research Scientific 
Computing Center supported by the Office
of Science of the DOE under Contract No. DE-AC02-05CH11231 and 
by the J\"ulich Supercomputing Center, awarded under the PRACE EU FP7 project 2011040546.

\end{document}

%% file: adjustment.tex
\usepackage{amsmath}

\newcommand{\brackets}[1]{\langle #1 \rangle}
\newcommand{\rbrackets}[1]{\left( #1 \right)}
\newcommand{\sbrackets}[1]{\left[ #1 \right]}

\newcommand{\epow}[1]{\mathrm{e}^{#1}}

\newcommand{\mpi}{m_{\pi}}

\newcommand{\fermi}{\,\mathrm{fm}}

\newcommand{\mev}{\,\mathrm{MeV}}

\newcommand{\ubar}{\bar{u}}

\newcommand{\balign}{\begin{align}}
\newcommand{\ealign}{\end{align}}
\newcommand{\beq}{\begin{equation}}
\newcommand{\eeq}{\end{equation}}
\newcommand{\balignat}[1]{\begin{alignat}{#1}}
\newcommand{\ealignat}{\end{alignat}}

\newcommand{\bfig}{\begin{figure}}
\newcommand{\efig}{\end{figure}}

\newcommand{\bc}{\begin{center}}
\newcommand{\ec}{\end{center}}

\newcommand{\btab}{\begin{table}}
\newcommand{\etab}{\end{table}}

\newcommand{\bcom}{}

\newcommand{\bitem}{\begin{itemize}}
\newcommand{\eitem}{\end{itemize}}

\newcommand{\benum}{\begin{enumerate}}
\newcommand{\eenum}{\end{enumerate}}

\newcommand{\qvec}{\vec{q}}
\newcommand{\Qvec}{\vec{Q}}
\newcommand{\xvec}{\vec{x}}
\newcommand{\yvec}{\vec{y}}

\newcommand{\matelem}{\mathcal{M}}


%% file: proceedings.bbl
\begin{thebibliography}{99}
\bibitem{Luscher:1985dn}
  M.~Luscher,
  Commun.\ Math.\ Phys.\  {\bf 104} (1986) 177.

\bibitem{Luscher:1986pf}
  M.~Luscher,
  Commun.\ Math.\ Phys.\  {\bf 105} (1986) 153.

\bibitem{Torok:2009dg}
  A.~Torok, S.~R.~Beane, W.~Detmold, T.~C.~Luu, K.~Orginos, A.~Parreno, M.~J.~Savage and A.~Walker-Loud,
  Phys.\ Rev.\ D {\bf 81} (2010) 074506
  [arXiv:0907.1913 [hep-lat]].

\bibitem{Feng:2010es}
  X.~Feng, K.~Jansen and D.~B.~Renner,
  Phys.\ Rev.\ D {\bf 83} (2011) 094505
  [arXiv:1011.5288 [hep-lat]].

\bibitem{Li:2007ey}
  X.~Li {\it et al.}  [CLQCD Collaboration],
  JHEP {\bf 0706} (2007) 053
  [hep-lat/0703015].

\bibitem{Lang:2012db}
  C.~B.~Lang and V.~Verduci,
  Phys.\ Rev.\ D {\bf 87} (2013) 5,  054502
  [arXiv:1212.5055].

\bibitem{McNeile:2002az}
  C.~McNeile {\it et al.}  [UKQCD Collaboration],
  Phys.\ Rev.\ D {\bf 65} (2002) 094505
  [hep-lat/0201006].

\bibitem{Alexandrou:2013ata}
  C.~Alexandrou, J.~W.~Negele, M.~Petschlies, A.~Strelchenko and A.~Tsapalis,
  Phys.\ Rev.\ D {\bf 88} (2013) 031501
  [arXiv:1305.6081 [hep-lat]].

\bibitem{Bernard:2001av}
  C.~W.~Bernard, T.~Burch, K.~Orginos, D.~Toussaint, T.~A.~DeGrand, C.~E.~Detar, S.~Datta and S.~A.~Gottlieb {\it et al.},
  Phys.\ Rev.\ D {\bf 64} (2001) 054506
  [hep-lat/0104002].

\bibitem{WalkerLoud:2008bp}
  A.~Walker-Loud, H.~-W.~Lin, D.~G.~Richards, R.~G.~Edwards, M.~Engelhardt, G.~T.~Fleming, P.~.Hagler and B.~Musch {\it et al.},
  Phys.\ Rev.\ D {\bf 79} (2009) 054502
  [arXiv:0806.4549 [hep-lat]].

\bibitem{Hemmert:1994ky}
  T.~R.~Hemmert, B.~R.~Holstein and N.~C.~Mukhopadhyay,
  Phys.\ Rev.\ D {\bf 51} (1995) 158
  [hep-ph/9409323].

\bibitem{Syritsyn:2007mp}
  S.~Syritsyn and J.~W.~Negele,
  PoS LAT {\bf 2007} (2007) 078
  [arXiv:0710.0425 [hep-lat]].

\bibitem{Beringer:1900zz}
  J.~Beringer {\it et al.}  [Particle Data Group Collaboration],
  Phys.\ Rev.\ D {\bf 86} (2012) 010001.

\end{thebibliography}
